\documentclass[reprint,amsmath,amssymb,aps,prb,groupedaddress,showpacs,floatfix,preprintnumbers]{revtex4-1}

\usepackage{graphicx,color}
\usepackage{subfigure}
\usepackage{verbatim}

\begin{document}
	\preprint{APS/123-QED}
	
	\title{Crystallographic and magnetic structure of UNi$_4$$^{11}$B}
	
	\author{J. Willwater$^1$, S. S\"ullow$^1$, M. Reehuis$^2$, R. Feyerherm$^2$, H. Amitsuka$^3$, B. Ouladdiaf$^4$, E. Suard$^4$, M. Klicpera$^5$, M. Vali\v{s}ka$^5$, J. Posp\'i\v{s}il$^5$ and V. Sechovsk\'y$^5$}
	
	\affiliation{$^1$Institut f\"ur Physik der Kondensierten Materie, TU Braunschweig, D-38106 Braunschweig, Germany\\
		$^2$Helmholtz-Center Berlin for Materials and Energy, D-14109 Berlin, Germany\\
		$^3$Department of Physics, Hokkaido University, Sapporo 060-0810, Japan\\
		$^4$Institute Laue-Langevin, BP 156, F-38042 Grenoble Cedex 9, France\\
		$^5$Faculty of Mathematics and Physics, Charles University, Ke Karlovu 5, Prague 2, Czech Republic}
	
	\date{\today}
	
	\begin{abstract}
		We present an extensive powder and single-crystal neutron scattering investigation of the crystallographic structure and magnetic order of the frustrated metallic $f$-electron magnet UNi$_4$B. We carry out a full refinement of the crystallographic structure and conclude that the low-temperature lattice symmetry is orthorhombic (space group \textit{Pmm}2; cell parameters: $a = 6.963(4)$\,\AA , $b = 14.793(9)$\,\AA , $c = 17.126(8)$\,\AA ). We determine the magnetically ordered structure, concluding that below $T_\mathrm{N} = 19.5~\mathrm{K}$ the material undergoes a transition into a partially ordered antiferromagnetic state. The magnetic structure is consistent with the existence of toroidal order in this material. We further test the proposal of a second magnetic transition occurring at $330~\mathrm{mK}$, concluding that the thermodynamic anomalies observed at these temperatures do not reflect modifications of the magnetic structure. Our study provides a consistent picture of the interrelationship of structural and magnetic properties in the frustrated magnet UNi$_4$B previously unresolved.     
	\end{abstract}
	
	\pacs{Valid PACS appear here}% PACS, the Physics and Astronomy
	% Classification Scheme.
	%\keywords{Suggested keywords}%Use showkeys class option if keyword
	%display desired
	\maketitle
	
	\section{\label{sec:level1}Introduction}
	
	Frustrated metallic $f$-electron magnets have been the focus of intense research efforts in recent years \cite{Vojta2018}. Examples are the frustrated (Kondo lattice) materials UNiGa \cite{Sechovsky1995}, CePdAl \cite{Doenni1996,Oyamada2008}, TbNiAl \cite{Maletta1998}, CeRhSn \cite{Schenck2004}, YbAgGe \cite{Schmiedeshoff2011} or CeIrSn \cite{Tsuda2018}. They all crystallize in the hexagonal ZrNiAl structure and the magnetic moments form a distorted Kagome lattice. This leads to a number of interesting effects like partial ordering in CePdAl \cite{Doenni1996} or complex magnetic phase diagrams with competing magnetic phases in UNiGa or YbAgGe \cite{Sechovsky1995,Schmiedeshoff2011}. Additionally, the susceptibilities for UNiGa, CePdAl and CeIrSn show a strong anisotropy between the measurements parallel and perpendicular to the kagome planes. From this an Ising-like magnetic behavior was proposed for these materials \cite{Sechovsky1995,Oyamada2008,Tsuda2018}.
	
	In the context of frustrated metallic $f$-electron magnets, $\mathrm{UNi_4B}$ has been considered as an early hexagonal example \cite{Mentink1994,Mentink1995,Mentink1996}. Mentink \textit{et al.} \cite{Mentink1994} reported the system to crystallize in the hexagonal $\mathrm{CeCo_4B}$ structure (space group: \textit{P}6/\textit{mmm}) with lattice parameters $a\nobreak=\nobreak4.953$\,\AA ~and $c\nobreak=\nobreak6.964$\,\AA . Here, the uranium ions form a triangular lattice in the hexagonal plane, which might enable magnetic frustration. 
	
	Thermodynamic studies and neutron scattering \cite{Mentink1994,Mentink1995,Mentink1996} revealed a magnetic transition at $T_\mathrm{N}\nobreak=\nobreak20~\mathrm{K}$. A Curie-Weiss temperature $\mathit{\Theta}_\mathrm{CW} \sim -65~\mathrm{K}$, significantly larger than the ordering temperature $T_\mathrm{N}$, indicates the strong antiferromagnetic interaction predominant in the system and has been taken to be indicative of magnetic frustration \cite{Mentink1994}. Based on these data, it was argued that the magnetic moments undergo a highly unusual form of partial antiferromagnetic ordering. Only two out of three magnetic moments should participate in long-range magnetic order, resulting in the formation of a vortex-like magnetic structure.  
	
	This interpretation raises several unresolved issues related to frustrated magnetism. First, the proposed magnetic structure does not reflect that in the \textit{P}6/\textit{mmm} space group there is just a single crystallographic site for the uranium ions, inconsistent with the inequivalence of this site in the magnetically ordered state. Therefore, it was argued that the formation of a crystallographic superstructure with a threefold enlargement of the in-plane lattice vector $a' = 3 a$ provides the structural inequivalence of the different uranium sites \cite{Mentink1996}. Correspondingly, the partial ordering was theoretically modeled as resulting from Kondo screening of the non-ordered uranium moments only \cite{Lacroix1996}. Moreover, low temperature specific heat measurements reveal a magnetic field dependent anomaly at $T^*\nobreak=\nobreak330~\mathrm{mK}$ \cite{Movshovich1999}. It was speculated that it signals a second magnetic phase transition where the remaining uranium ions undergo an ordering transition. Attempts to verify this proposal have been unsuccessful so far. Very recently, it has been proposed that the feature in the specific heat represents a Schottky anomaly from a crystal field splitting possible in the orthorhombic symmetry \cite{Yanagisawa2021}.
	
	Pulsed field magnetization data reveal large anisotropy between the measurement parallel and perpendicular to the hexagonal plane ($ab$ plane) \cite{Mentink1995,Mentink1996}. The measurements for $B||a$ and $B||b$ show multiple steps in the magnetization. To analyze these features, the magnetization was simulated using a quasi-one-dimensional XY model. The results were interpreted to the effect, that the small magnetization steps are triggered by a slight reorientation of the above described magnetic structure and that a jump in the magnetization at higher fields is due to an alignment of the magnetic moments along the external magnetic field \cite{Mentink1995,Mentink1996}.
	
	More recently, the magnetic structure of $\mathrm{UNi_4B}$ was discussed in the context of toroidal order \cite{Saito2018}. It arises in states with complex magnetic moment configurations in the local absence of spatial inversion and time reversal symmetry \cite{Hayami2014}. In toroidally ordered systems novel types of magnetoelectric effects might occur. The proposed vortex-like ordering of the magnetic moments in $\mathrm{UNi_4B}$ is equivalent to the theoretically investigated toroidal ordering in Ref. \cite{Hayami2014}. Therefore, $\mathrm{UNi_4B}$ was discussed as a metallic example of a system exhibiting toroidal order \cite{Hayami2014}. As well, it was argued that this could explain the current dependent magnetization observed in $\mathrm{UNi_4B}$ \cite{Saito2018}. 
	
	Such an interpretation relies on the correct characterization of the crystallographic and correspondingly the magnetic symmetry. Recently, however, X-ray diffraction \cite{Haga2008} and NMR \cite{Takeuchi2020} data have indicated that the crystallographic symmetry of $\mathrm{UNi_4B}$ is lower than the proposed hexagonal one. From the analysis of single-crystal X-ray diffraction data, the best description of the crystal structure at $300~\mathrm{K}$ was reported using the space group 63 (\textit{Cmcm}). More recently, an independent study of the crystallographic structure by means of synchrotron X-ray diffraction supports this notion of an orthorhombic lattice, also with the conclusion of a lattice symmetry with space group \textit{Cmcm} \cite{tabata2021}.
	
	In this situation, we have carried out neutron diffraction experiments on powder and single crystalline samples to characterize the crystallographic structure and revisit the issue of the magnetically ordered structure of $\mathrm{UNi_4B}$. A study of the crystallographic structure of UNi$_4$B by neutron scattering allows to utilize the much brighter elemental contrast between U, Ni and B, as compared to a synchrotron X-ray investigation. The neutron scattering cross sections for the different elements (from Ref. \cite{Sears1992}) are U: $8.908~\mathrm{barn}$, Ni: $18.5~\mathrm{barn}$ and $^{11}$B: $5.77~\mathrm{barn}$, while for X-ray diffraction they scale with the square of the atomic number $Z$ of the elements. Thus, while synchrotron studies typically yield a higher signal-to-noise ratio for the raw experimental data than neutron scattering, because of the different and often brighter elemental contrast neutron scattering provides a complementary view of the structural properties of a material. This is in particular effective for materials with a large difference of atomic number of the different elements, as is regularly demonstrated in studies of materials containing for instance hydrogen \cite{Heinze2021} (in our case, the elemental contrast between boron and uranium is more than a factor of 200 brighter in neutron scattering than in X-ray diffraction). In addition, we use neutron scattering to investigate the magnetic behavior of $\mathrm{UNi_4B}$ below $T_\mathrm{N}$ and at temperatures around the proposed second phase transition at $T^*$.
	
	\section{\label{sec:level2}Experimental details}
	
	A high-quality UNi$_4$$^{11}$B single crystal has been prepared by the floating zone method which was found recently very effective for the growth of uranium materials \cite{Honda2019,Tabata2017}. A commercial four-mirror optical furnace with halogen lamps each 1\,kW (model FZ-T-4000-VPM-PC, Crystal Systems Corp., Japan) was used. In the first step, a polycrystalline material of UNi$_4$$^{11}$B was synthesized by arc-melting from the stoichiometric amounts of pure elements U (3N, further treated by Solid-State Electrotransport \cite{Haga1998,Pospisiil2011}), Ni (4N), and $^{11}$B isotope in Ar (6N) protective atmosphere. Any sign of evaporation of the components was not detected during the melting. Then, a precursor in the form of a 40\,mm long rod of diameter 6\,mm was prepared by arc melting in a special water-cooled copper mold at identical protective conditions. The quartz chamber of the optical furnace was evacuated by a turbomolecular pump to $10^{-6}$\,mbar before the crystal growth process. To desorb gases from the surface of the precursor, the power of the furnace was increased gradually up to 15\,\% of maximum power (far from the melting $\sim$30\,\% and growing power $\sim$35\,\%) and the precursor was several times passed through the hot zone while continuously evacuating. The whole growth process was also performed with continuous evacuation and vacuum $10^{-6}$\,mbar. A narrow neck was created at the beginning of the growth process by variation of the speed of the upper and bottom pulling shafts. The pulling rate was very slow at only 0.6\,mm/h without rotation. A single crystal of cylindrical shape of length $\sim$60 mm and varying diameter 3-5 mm was obtained. The high quality and orientation of the single crystal were verified by the X-ray and neutron Laue method. The chemical composition of the single crystal was verified by scanning electron microscopy (SEM) using a Tescan Mira I LMH system equipped with an energy-dispersive X-ray detector (EDX) Bruker AXS. The analysis revealed a single crystal of the expected nominal composition. The ingot was cut by a fine wire saw to roughly a cube sample for single-crystal neutron scattering experiments. The rest of the ingot was ground into a fine powder ($8$\,g) and used for the powder neutron diffraction presented in this work.
	
	The powder neutron diffraction experiments were performed at the D2B instrument (ILL Grenoble) with a neutron wavelength of $1.594$\,\AA ~at temperatures of $2$\,K, $30$\,K and $300$\,K. The CYCLOPS Laue diffractometer was used for both the Laue characterization of the investigated single crystal and for low-temperature studies of the magnetic signal down to $40$\,mK ($^3\mathrm{He}$/$ ^4\mathrm{He}$ dilution cryostat) \cite{Ouladdiaf2011}. Additionally, two experiments were carried out at the four-circle diffractometer D10 at the ILL Grenoble. Around 4500 nuclear (1300  independent) reflections at temperatures of $2$\,K and $30$\,K and 311 magnetic reflections from which 129 had intensity (15 unique) at $2$\,K were measured  using wavelengths of $\mathit{\lambda}~=~1.26$\,\AA ~and $\mathit{\lambda}~=~2.36$\,\AA. The measurements continued at the latter wavelength with the crystal cooled in a $^3\mathrm{He}$/$ ^4\mathrm{He}$ dilution cryostat down to $100$\,mK. The data were analyzed using the FullProf package \cite{Rodriguez-Carvajal1993} and can be accessed at Refs. \cite{ILL,ILL2,ILL3}.
	
	\section{\label{sec:level3}Results and discussion}
	
	\subsection{Crystallographic structure of UNi$\mathbf{_4}$B}
	
	\begin{figure}[t]
		\includegraphics[scale=0.105]{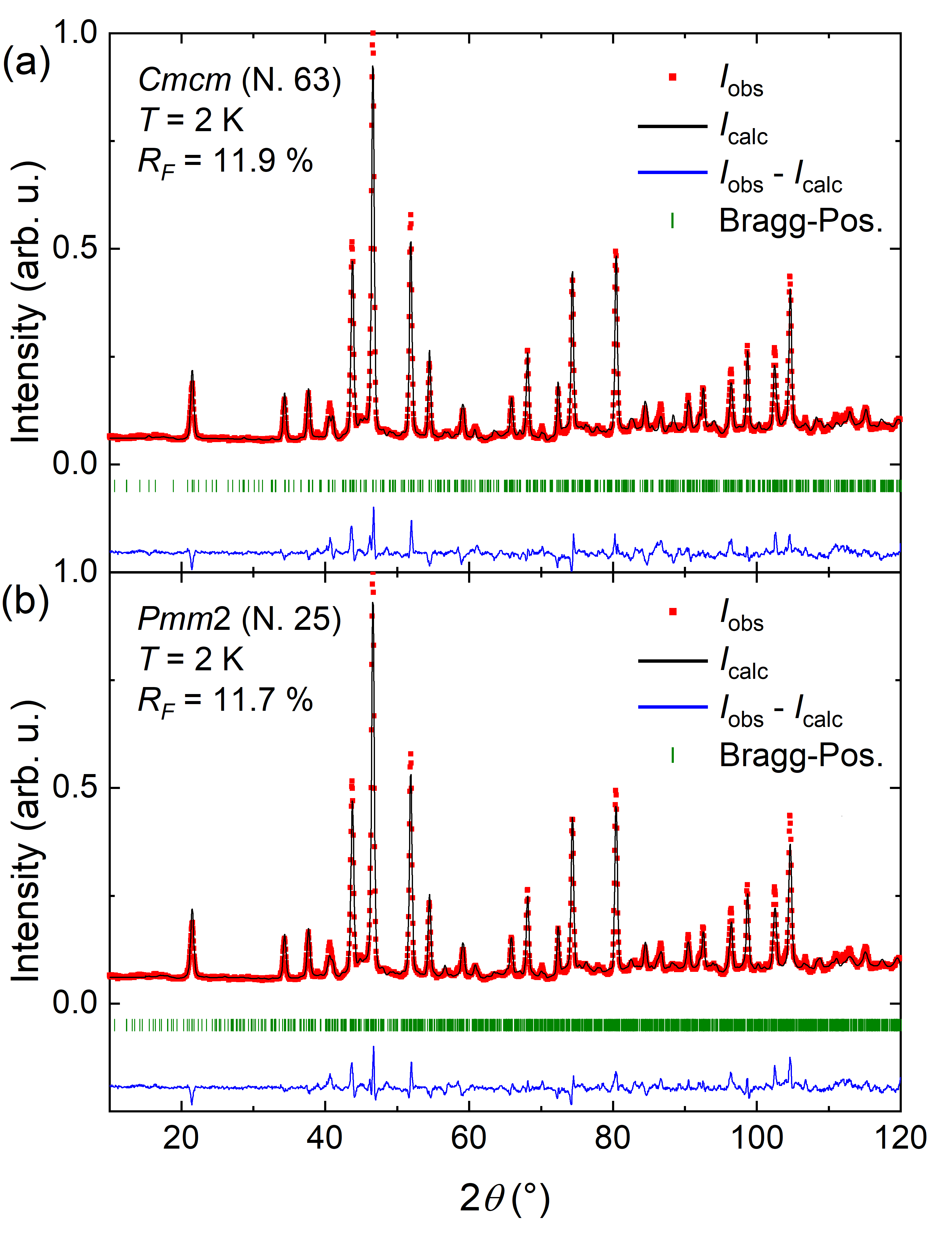}
		\caption{\label{powder} (Color online) Structural refinement of $\mathrm{UNi_4B}$ from powder neutron diffraction data with the space group (a) \textit{Cmcm} and (b) \textit{Pmm}2; for details see text.}
	\end{figure}
	
	We started our study with the neutron diffraction experiment on the $\mathrm{UNi_4B}$ powder sample (see Fig. \ref{powder} for the measurement at $2~\mathrm{K}$). A refinement with the original hexagonal structure (space group \textit{P}6/\textit{mmm}) \cite{Mentink1994} yields a poor fitting result. Therefore, as a starting point the description of the structure with the orthorhombic space group \textit{Cmcm} (number: 63) from Ref. \cite{Haga2008} was chosen and the software FullProf was used for the refinement (s. Fig. \ref{powder} (a)). We obtain a basic description of the lattice symmetry with the space group \textit{Cmcm} (cell parameter: $a\nobreak=\nobreak6.963(7)$\,\AA , $b\nobreak=\nobreak17.088(3)$\,\AA , $c\nobreak=\nobreak14.825(8)$\,\AA ). This is in a good agreement with the reported lattice parameters from room temperature (synchrotron) X-ray diffraction in Ref. \cite{Haga2008} ($a\nobreak=\nobreak6.968$\,\AA , $b\nobreak=\nobreak17.1377$\,\AA , $c\nobreak=\nobreak14.8882$\,\AA ) and Ref. \cite{tabata2021} ($a\nobreak=\nobreak6.9395(1)$\,\AA , $b\nobreak=\nobreak17.0524(3)$\,\AA , $c\nobreak=\nobreak14.8063(7)$\,\AA ). In detail, however, there are deviations between data and refinement. For example, as can be observed in the difference pattern $I_{obs} - I_{calc}$, the intensities of the strong peaks are not very well described by the fit. Also, for larger scattering angles the refinement does not reproduce the data in full detail. This is reflected by an only moderately low refinement factor 
	\begin{align}
		R_F\nobreak=\nobreak \frac{\sum_{k} |F_{\mathrm{obs},k}-F_{\mathrm{calc},k}|}{\sum_{k} |F_{\mathrm{obs},k}|} \nobreak=\nobreak11.9\%
	\end{align}
	with $F_{\mathrm{obs}}$ ($F_{\mathrm{calc}}$) the measured (calculated) structure factor. Similar results were obtained for the measurements at $30~\mathrm{K}$ and $300~\mathrm{K}$. 
		
	\begin{figure}[t]
		\includegraphics[scale=0.32]{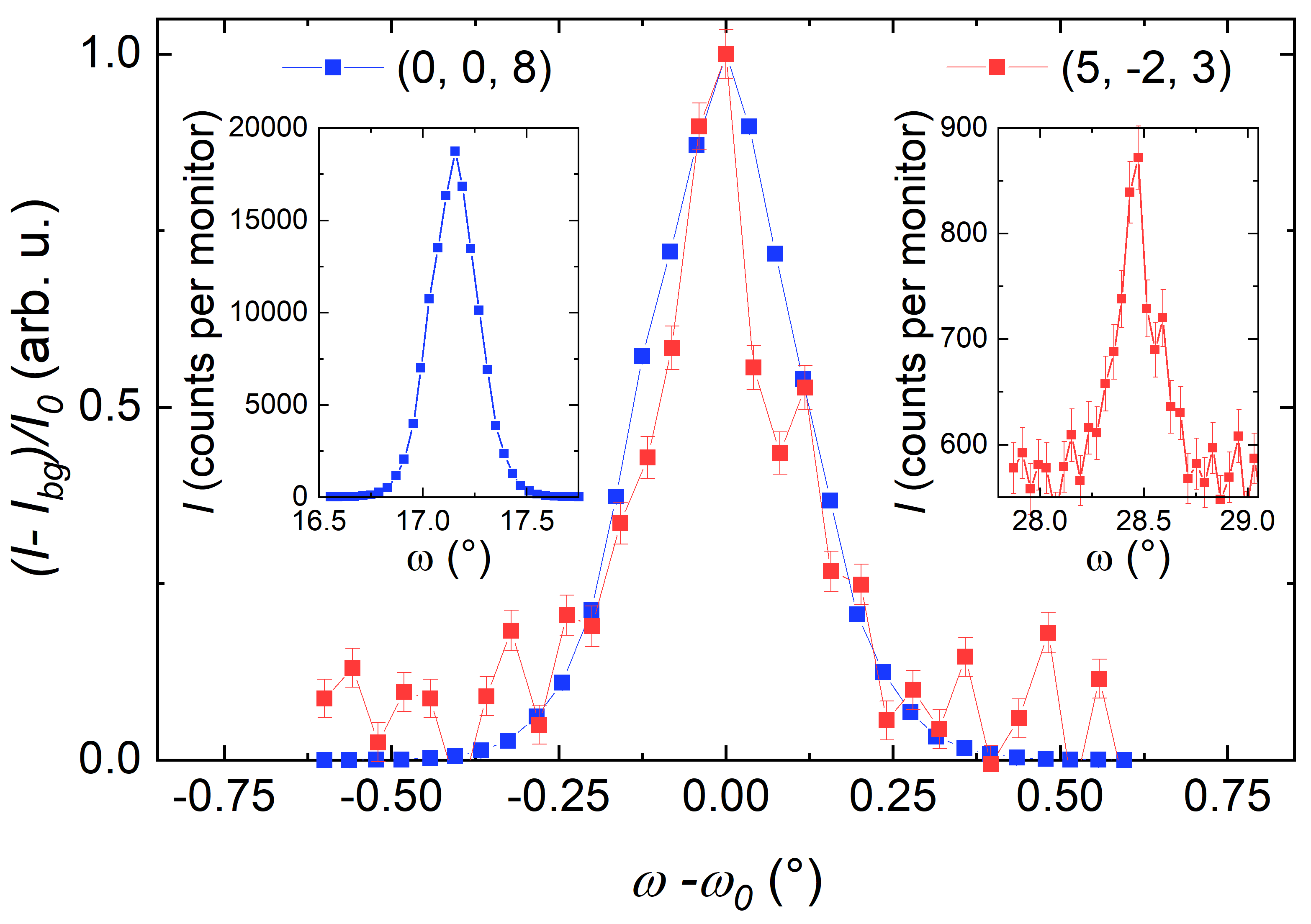}
		\caption{\label{BPC} (Color online) $\omega$-scans of two structural Bragg peaks of UNi$_4$B: the $(0, 0, 8)$ peak represents one with almost maximum intensity allowed for the \textit{Cmcm} space group, the $(5, -2, 3)$ peak is a low intensity peak violating the \textit{Cmcm} scattering condition, $hkl:\nobreak h+k \nobreak=\nobreak2n$; insets depict the original data of the peaks, including experimental background.}
	\end{figure}
		
	Even with the orthorhombic structural model there is substantial room for improvement of the refinement. To optimize the crystal structure analysis of $\mathrm{UNi_4B}$, we have carried out neutron diffraction experiments on single crystalline material. Surprisingly, we measured a number of low intensity (up to 0.1\,\% of maximum intensity) reflections that violate the conditions $hkl:\nobreak h+k \nobreak=\nobreak2n$ and $00l:\nobreak l\nobreak=\nobreak2n$ of the \textit{Cmcm} space group. As an example, in Fig. \ref{BPC} we plot the normalized peak intensity for the \textit{Cmcm} forbidden $(5, -2, 3)$ Bragg peak in an $\omega$-scan (original data see inset Fig. \ref{BPC}). A comparison with the $(0, 0, 8)$ peak (about a factor of 40 more intense) illustrates that the forbidden Bragg peaks are not broadened in comparison to the allowed Bragg peaks (Fig. \ref{BPC}). The $hkl:\nobreak h+k \nobreak=\nobreak2n$ condition is given by the \textit{C}-centering and since this condition has been violated, it can be concluded that the \textit{C}-centering is lost. In additional experiments we have verified that $\mathit{\lambda}/2$ contamination can be ignored. Therefore, we conclude that the single crystal $\mathrm{UNi_4B}$ studied in this work is not crystallizing in the \textit{Cmcm} space group, but in a lattice of lower symmetry.
	
	\begin{figure}[t]
		\includegraphics[scale=0.37]{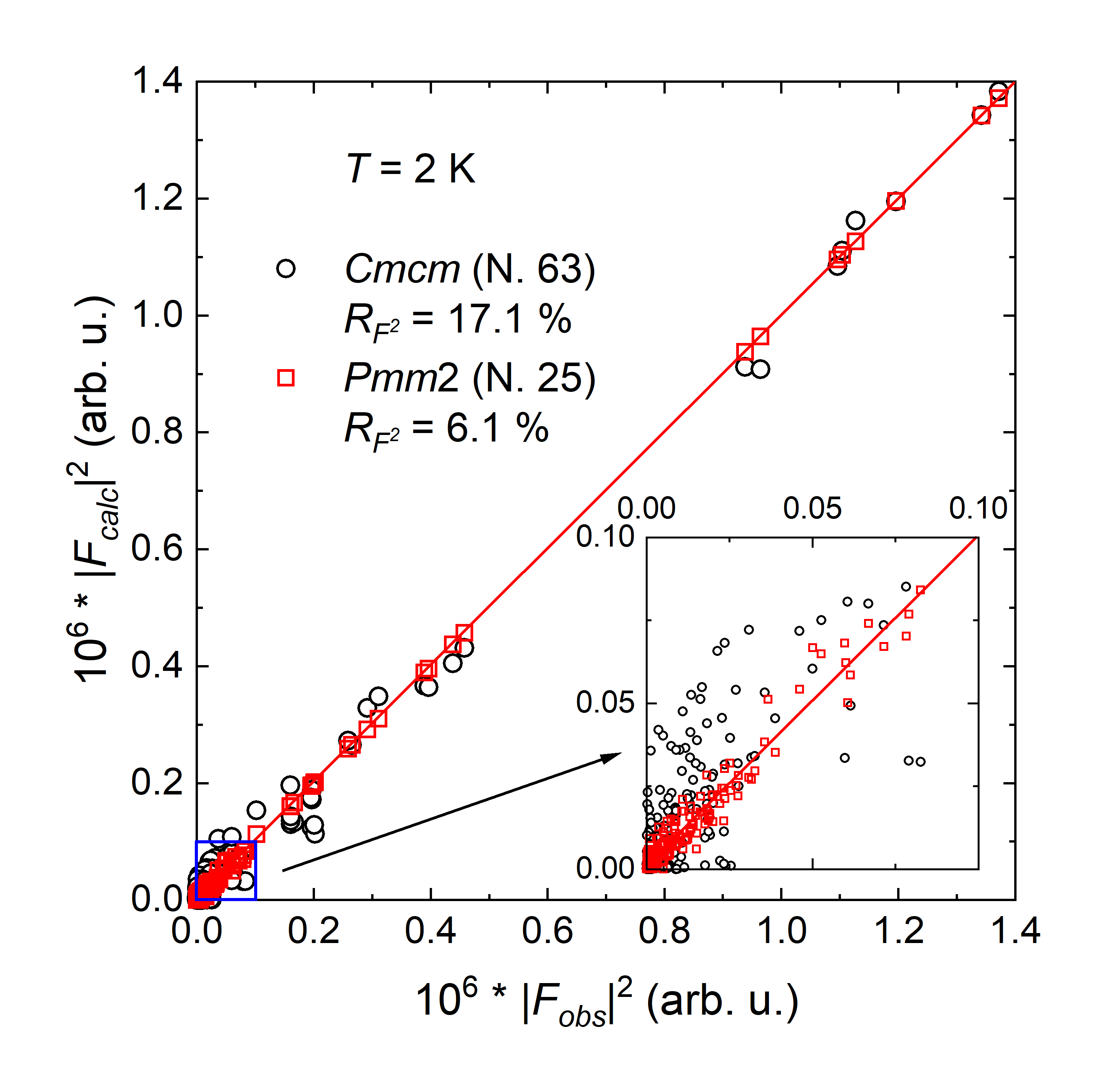}
		\caption{\label{SC} (Color online) Structural refinements of $\mathrm{UNi_4B}$ from single crystal neutron diffraction data ($\mathit{\lambda}~=~2.36$\,\AA) with the space groups \textit{Cmcm} and \textit{Pmm}2; for details see text.}
	\end{figure}

	To account for these observations, we have tested different related space groups of next-lower symmetry: \textit{Pbcm} (number: 57), \textit{Pmc}2$_1$ (26), \textit{Pmm}2 (25) and \textit{P}222$_1$ (17). By comparing the indices of our measured reflections with the diffraction conditions for a given space group only the orthorhombic space group \textit{Pmm}2 (25) satisfies the extinction rules. In addition, the best structural description by refining the atomic positions and the isotropic displacement parameters was obtained with \textit{Pmm}2 (cell parameters: $a = 6.963(4)$\,\AA , $b = 14.793(9)$\,\AA ~and $c = 17.126(8)$\,\AA ; for a full set of atomic positions see \cite{SuppMat}).
	
	\begin{figure}[t]
		\includegraphics[scale=0.43]{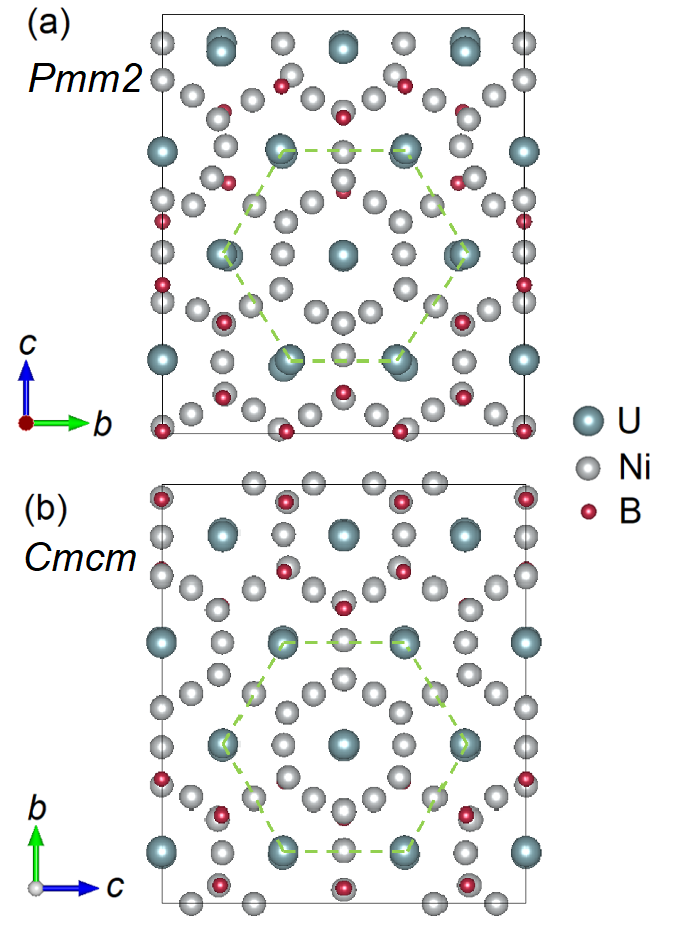}
		\caption{\label{str_vgl} (Color online) Crystal structure of $\mathrm{UNi_4B}$ with the space group (a) \textit{Pmm}2 (b) \textit{Cmcm} (shifted by a translation operator for better comparison); for details see text.}
	\end{figure}
	
	While the quality of the refinement of the powder neutron diffraction data with space group \textit{Pmm}2 is similar to the one with space group \textit{Cmcm} [$R_F\nobreak=\nobreak11.7$\,\%; see Fig. \ref{powder} (b)] \cite{powder}, there are clear quality differences in the single crystal refinement. In Fig. \ref{SC} we compare the results of structural refinements, using both the space groups 63 (\textit{Cmcm}) and 25 (\textit{Pmm}2). To carry out these refinements, the reflections with finite intensity were measured, each of the peaks was fitted with a pseudo-Voigt function and the intensities were obtained by integrating these peak functions. In addition, an absorption correction was applied. In Fig. \ref{SC} we document the resulting observed intensities vs. the calculated ones. Clearly, the refinement of the structure with the space group \textit{Pmm}2 shows a better agreement with the measured intensities than space group \textit{Cmcm}. In particular, for the low intensity peaks the deviations of the calculated intensities from the measured ones are significantly larger for the space group \textit{Cmcm} than for \textit{Pmm}2 (see inset of Fig. \ref{SC}). This is also validated by a lower $R$-factor for \textit{Pmm}2, $R_{F^2}\nobreak=\nobreak6.1$\,\%, in comparison to $R_{F^2}\nobreak=\nobreak17.1$\,\% for \textit{Cmcm} (for more details see \cite{SuppMat}). The $R$-factors for the other candidate space groups listed above turned out to be significantly larger as well, with values $R_{F^2} > 20$\,\%. Similar results were obtained at $30~\mathrm{K}$ and in the second experiment with a neutron wavelength of $1.26$\,\AA . Thus, the space group \textit{Pmm}2 produces the optimum refinement of our single-crystal neutron diffraction data. From subsequent fits of the powder neutron diffraction data using this space group, we obtain the lattice parameters of $\mathrm{UNi_4B}$ from low to room temperature listed in Tab. \ref{lattice}.
	
	With the conclusions from our neutron diffraction experiments we are now able to characterize the crystal structure of $\mathrm{UNi_4B}$. In Fig. \ref{str_vgl} we compare the refined crystal structures \textit{Pmm}2 and \textit{Cmcm} with a view of the quasi-hexagonal plane for both space groups. From the figure, the close resemblance of the two structures is apparent (note that the lattice parameter labellings of $b$ and $c$ axis have been interchanged between \textit{Pmm}2 and \textit{Cmcm}; see below). Especially, the close-to-hexagonal arrangement of the uranium atoms is visible for both lattices and therefore, magnetic frustration is a relevant factor in the physics of $\mathrm{UNi_4B}$.

	\begin{table}
		\caption{\label{lattice} Lattice parameters of $\mathrm{UNi_4B}$ as function of temperature, derived from the structural refinement with the space group \textit{Pmm}2 of neutron powder diffraction data; for details see text.}
		\begin{ruledtabular}
			\begin{tabular}{llll}
				$T$ (K) & $a$ (\AA ) & $b$ (\AA ) & $c$ (\AA )\\ \hline
				2 & 6.963(4) & 14.793(9) & 17.126(8) \\
				30 & 6.963(4) & 14.795(3) & 17.127(1) \\
				300 & 6.973(6) & 14.828(8) & 17.157(9) \\
			\end{tabular}
		\end{ruledtabular}		
	\end{table}
	
	In detail, some differences can be seen in the positions of the various atoms. To understand these differences, it is instructive to relate the atomic coordinates in the transformation \textit{Pmm}2 $ \leftrightarrow$ \textit{Cmcm}. The \textit{Pmm}2 lattice is derived from \textit{Cmcm} by relaxing symmetry conditions for atomic coordinates and thus reducing the multiplicities of these sites. As an example, the Ni atoms labeled Ni$_{3\text{-}1}$ to Ni$_{3\text{-}4}$ in the \textit{Pmm}2 lattice (for more details see \cite{SuppMat}) are derived from a single atomic position in the \textit{Cmcm} structure, {\it i.e.}, Ni$_{3}$ with coordinate (0, 0.28049, 1/4). The transformation \textit{Cmcm} $\leftrightarrow$ \textit{Pmm}2 involves the interchange $b \leftrightarrow c$ axis and a shift of the unit cell basis by 0.25 along $c$. Then, the $b$ coordinate of Ni$_{3}$, 0.28049, which with the symmetry operations in \textit{Cmcm} defines four Ni positions, becomes an independent parameter for Ni$_{3\text{-}1}$ to Ni$_{3\text{-}4}$. In our refinement, we obtain values for the $c$ coordinate of 0.26358 (Ni$_{3\text{-}1}$), 0.84428 (Ni$_{3\text{-}2}$), 0.76943 (Ni$_{3\text{-}3}$) and 0.31426 (Ni$_{3\text{-}4}$), implying that these are individual atomic positions close to the Ni$_{3}$ sites of \textit{Cmcm}. In other words, an atomic position in the \textit{Cmcm} lattice is replaced by four closely related, but symmetry lowered atomic positions in the \textit{Pmm}2 lattice. 
	
	To summarize, the main differences between the two descriptions of the crystal structure are the positions of the Ni and B atoms. Therefore, to correctly describe the crystallographic structure of UNi$_4$B it is necessary to lower the symmetry to \textit{Pmm}2. Conversely, the structural arrangement of the U atoms is hardly affected by the transformation \textit{Pmm}2 $\leftrightarrow$ \textit{Cmcm}. As a consequence, based on X-ray diffraction experiments the \textit{Cmcm} structure was reported, as X-ray diffraction is much more sensitive to atoms with high atomic numbers. It is likely that the lower structural symmetry has an impact on the magnetic behavior of UNi$_4$B. The magnetically inequivalent U atoms are now described by different structural positions and it may be assumed that the distortion of the Ni-B sublattice in the hexagonal plane affects the canting of the magnetic moments.
		
	\begin{figure}[t]
		\includegraphics[scale=0.48]{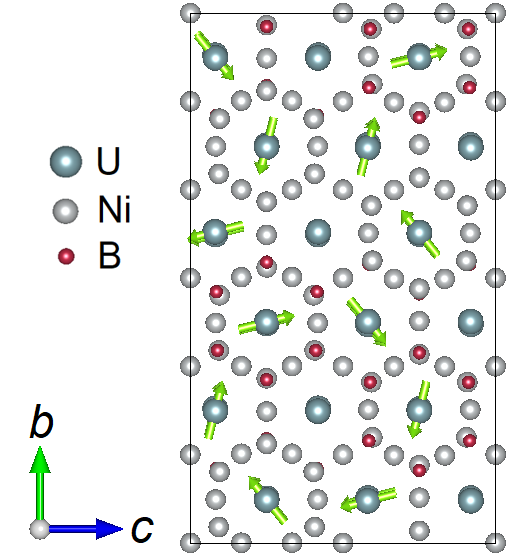}
		\caption{\label{str_magn} (Color online) Magnetic unit cell with the proposed magnetic structure; for details see text.}
	\end{figure}

	It should be mentioned that for some weak reflections ($F^2_{\mathrm{clac}}\nobreak<\nobreak1$\,\% of the maximal $F^2_{\mathrm{clac}}$) the intensity calculated with the space group \textit{Pmm}2 overestimates the measured intensity. This could indicate that there still is an additional reflection condition ($0kl:\nobreak k\nobreak=\nobreak2n$) and that UNi$_4$B might possibly be described by a space group of slightly higher symmetry. However, attempts to refine the experimental data with a lattice of symmetry between \textit{Pmm}2 and \textit{Cmcm} have produced refinement factors larger than for \textit{Pmm}2 in this work. Based on the experimental data, we thus conclude that the optimum solution for the crystal structure of UNi$_4$B is the orthorhombic \textit{Pmm}2 lattice. 
	
	This slight ambiguity regarding the "intrinsic" lattice symmetry of UNi$_4$B might be considered also from a different angle. Inevitably, given the large number of free parameters in our \textit{Pmm}2 refinement, parameter interdependency has to be considered carefully. Moreover, given the fundamental structural similarity of the \textit{Pmm}2 and \textit{Cmcm} refined structures, the question arises if the symmetry-lowered atomic positions in the \textit{Pmm}2 refinement reflect local distortions arising for instance from strain. Such behavior, however, would be more appropriately captured as a random distribution of atomic positions around the nominal atomic coordinate by careful refinement of the displacement parameters (as an example, see the case of UPt$_2$Si$_2$ \cite{Sullow2008,Prokes2020}). A study of the neutron pair distribution function could provide more information about the structural disorder. The experimental fact that we observe distinct structural Bragg peaks which are symmetry-forbidden in the \textit{Cmcm} structure validates our approach of a crystal structure refinement based on the \textit{Pmm}2 symmetry. For all practical purposes of subsequently refining the magnetic structure of UNi$_4$B, these structural details will be of no relevance. 
	
	\subsection{Magnetic order in UNi$_\mathbf{4}$B}

	\begin{figure}[t]
		\includegraphics[scale=0.1]{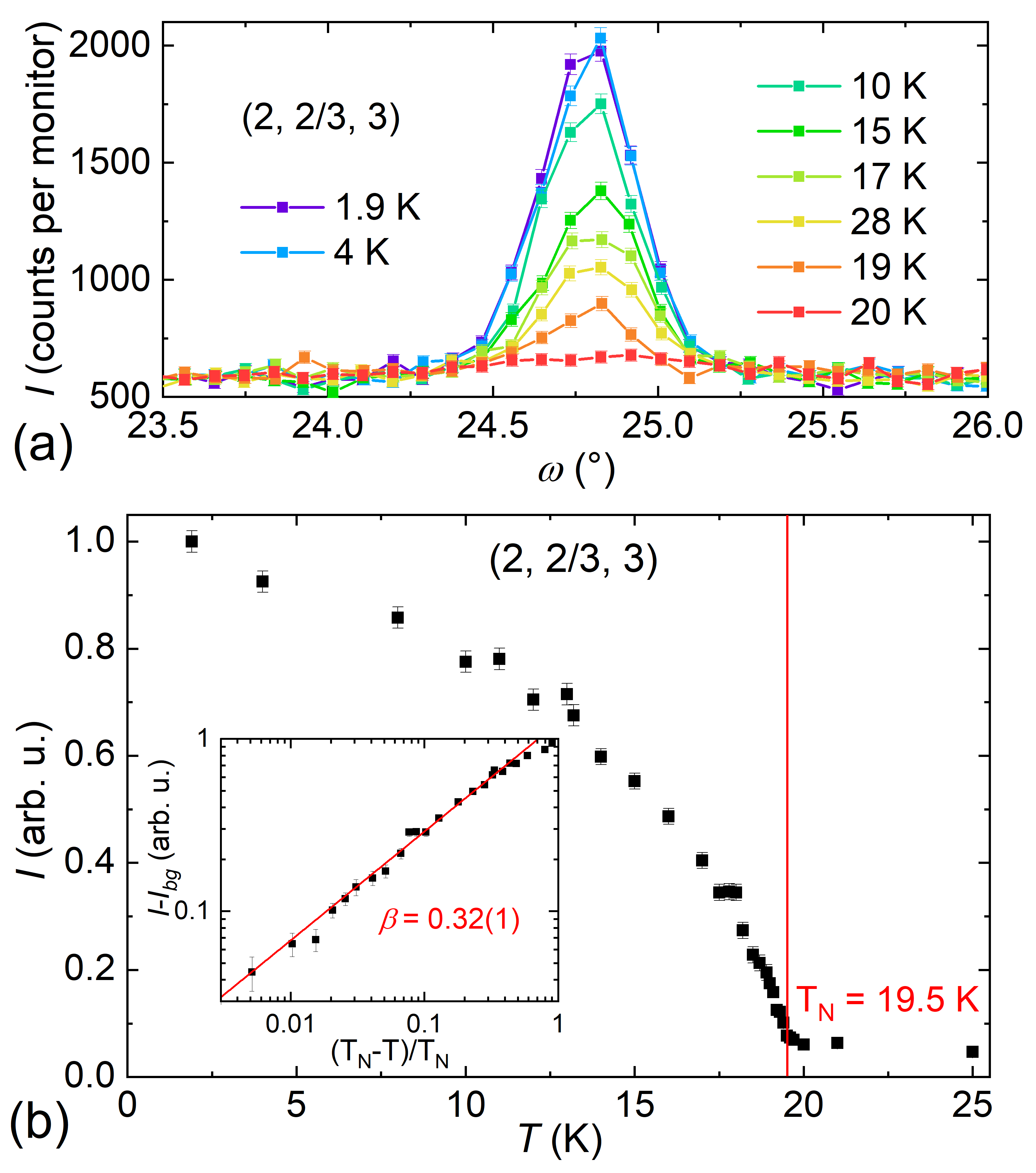}
		\caption{\label{tempdep} (Color online) (a) $\omega$-scans of the magnetic (2,\,2/3,\,3) Bragg peak and (b) temperature dependence of the integrated intensities of the magnetic reflection at (2,\,2/3,\,3). Inset: Double logarithmic plot of the integrated intensity vs. $(T_\mathrm{N}-T)/T_\mathrm{N}$ to illustrate the critical behavior; for details see text.}
	\end{figure}
	
	Due to the complexity of the $\mathrm{UNi_4B}$ crystal structure description within the space group \textit{Pmm}2, there are sixteen crystallographically inequivalent uranium positions. Therefore, with the close structural similarity of the \textit{Pmm}2 and \textit{Cmcm} lattices, for the investigation of the magnetically ordered phase we will continue with a structural description using the space group \textit{Cmcm}. The positions of the uranium atoms change only slightly between \textit{Pmm}2 and \textit{Cmcm}, while the description in the higher symmetry requires less atom positions (18 in \textit{Cmcm} vs. 68 in \textit{Pmm}2).
	
	The best solution for the refinement of the magnetic structure was obtained with a propagation vector of $\mathbf{k}\nobreak=\nobreak(0,2/3,0)$, leading to a magnetic unit cell with the dimensions $a\times 1.5b\times c$ based on the \textit{Cmcm} symmetry. This magnetic structure implies the existence of uranium atoms that do not carry ordered magnetic moments. The refinement was started with the magnetic structure proposed by Mentink \textit{et al.} \cite{Mentink1995}. Here, the magnetic moments form a vortex-like structure with an angle of $120^\circ$ to the next magnetic moment in the vortex. We have recorded a total of 311 magnetic reflections which could be generated by $(hkl)_M\nobreak=\nobreak(hkl)_N \pm \mathbf{k}$. A closer inspection showed that only nuclear reflections with $h\nobreak=\nobreak2n$ and $k\nobreak=\nobreak2n$ generate magnetic intensities. Thus, a total of 129 (15 unique) magnetic reflections were used for the refinement. In the final refinement the calculated intensities were in a good agreement with the intensities of the merged reflections which resulted in the residuals $R_{F^2}\nobreak=\nobreak8.48$\,\% and $R_{F}\nobreak=\nobreak7.35$\,\%. In Ref. \cite{Mentink1995}, where only 9 unique magnetic reflections were used, a somewhat larger residual $R_{I}\nobreak=\nobreak11.4$\,\% was obtained.
		
	Next, from our refinement we obtain an ordered magnetic moment $\mu_{ord}\nobreak=\nobreak 0.99(1)~\mu_\mathrm{B}$/(U atom). This value is somewhat smaller than reported in Ref.  \cite{Mentink1995}, $\mu_{\mathrm{ord}}\nobreak=\nobreak 1.2(2)~\mu_\mathrm{B}$/(U atom). In our case,  the value has been derived from a full refinement of the magnetic structure, rather than comparing the intensity of a magnetic and a structural Bragg peak in Ref.  \cite{Mentink1995}, allowing a more accurate moment determination in our study. The moment value is in good agreement with the saturation magnetization observed in high-field magnetization measurements \cite{Mentink1996}.
	
	\begin{figure}[t]
		\includegraphics[scale=0.12]{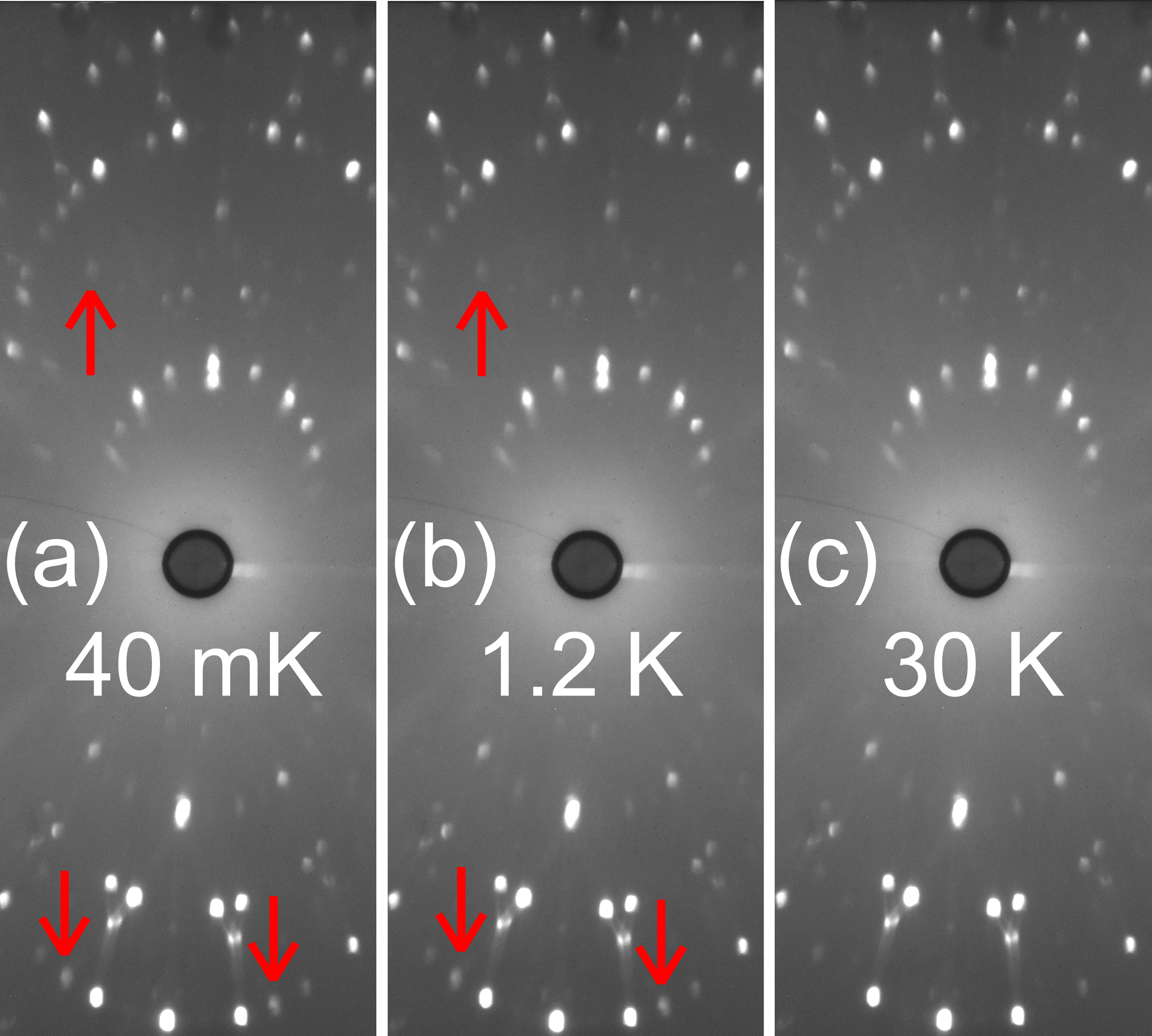}
		\caption{\label{cyclops} Section of a neutron Laue picture from $\mathrm{UNi_4B}$ at (a) $0.04$\,K, (b) $1.2$\,K and (c) $30$\,K. Some magnetic reflections are marked with arrows; for details see text.}
	\end{figure}

	The result of the refinement of the magnetic structure can be improved, if a slight canting of the magnetic moments is allowed. The best result of our refinement with $R_{F^2}\nobreak=\nobreak7.06$\,\% and $R_{F}\nobreak=\nobreak5.09$\,\% is displayed in Fig. \ref{str_magn}. According to this refinement, within the plane of the vortex the magnetic moments are canted up to $15^\circ$ in comparison to the perfect vortex structure. Contrary, the magnetic moments seem not to be canted out of the $bc$ plane. Taking the previous crystal structure analysis into account, it could be argued that the slight distortions of the nickel and boron atomic positions in the $bc$ plane lead to a canting of the magnetic moments. Of course, due to the large number of free parameters in the refinement, this observation will require further experiments to firmly prove it. Altogether, our study fully establishes partial ordering and the magnetic vortex structure in orthorhombic $\mathrm{UNi_4B}$.
	
	\begin{figure}[t]
		\includegraphics[scale=0.35]{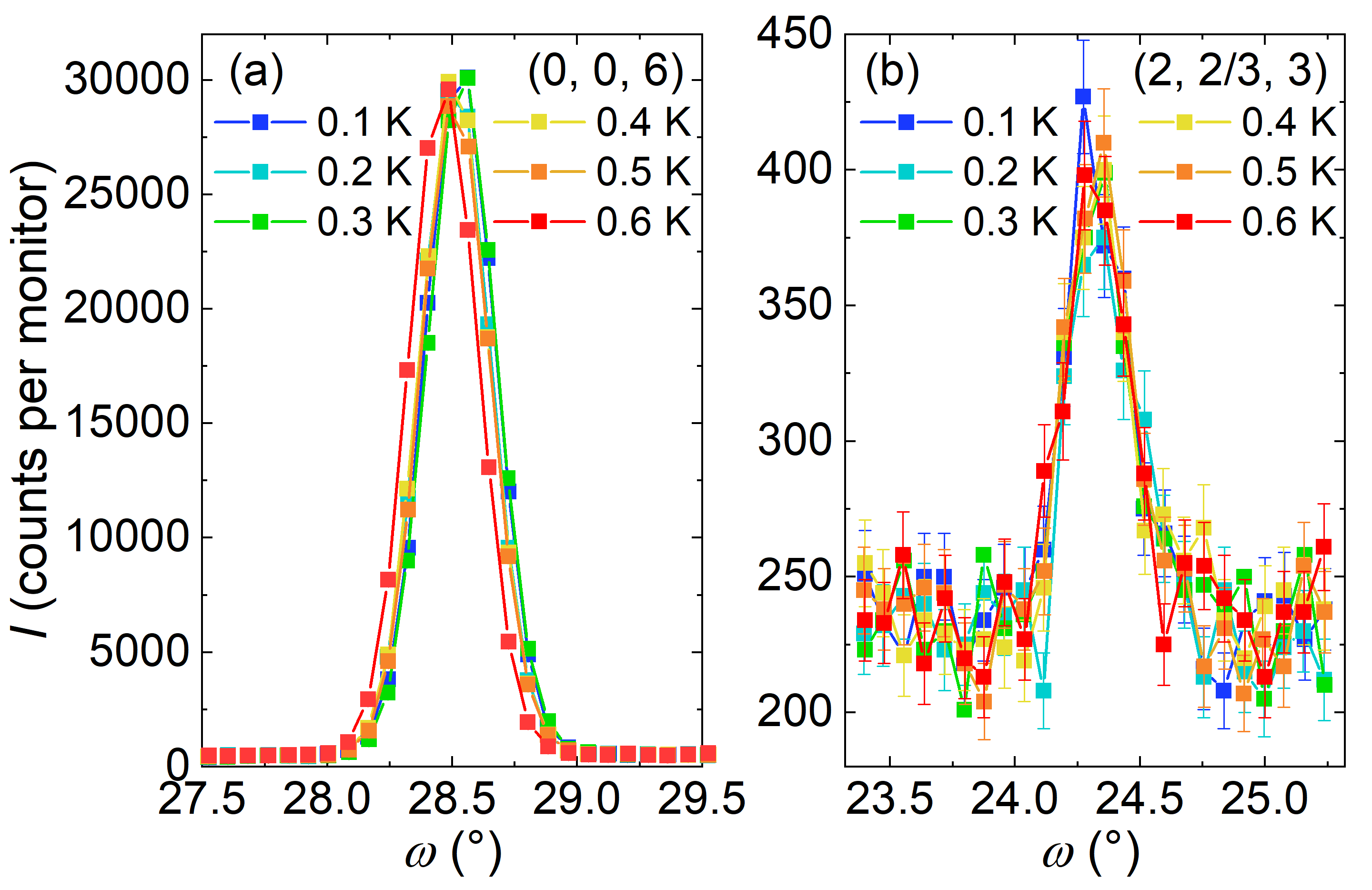}
		\caption{\label{magn_low} (Color online) $\omega$-scans around the (a) structural (0,\,0,\,6) and (b) magnetic (2,\,2/3,\,3) Bragg peaks at low temperatures; for details see text.}
	\end{figure}
	
	To improve the analysis of the magnetic structure of $\mathrm{UNi_4B}$, in principle one might carry out experiments using local probes such as NMR or $\mu$SR. Conceptually, provided the hyperfine interactions in a given material are known, it is possible to associate a specific spin arrangement to local magnetic fields and field distributions seen in NMR or $\mu$SR studies (see for instance the case of the 115-superconductors \cite{Curro2006}). Only, local magnetic field distributions detected for instance in $^{11}$B-NMR or $\mu$SR experiments on $\mathrm{UNi_4B}$ are rather unspecific \cite{Takeuchi2020,Nieuwenhuys1995}. At present, it seems unlikely that more knowledge on the magnetic structure of $\mathrm{UNi_4B}$ can be obtained along this route.
	
	The temperature dependence of the intensity of the magnetic reflections was investigated by carrying out $\omega$-scans at different temperatures as shown for the magnetic (2,\,2/3,\,3) Bragg peak in Fig. \ref{tempdep} (a). The magnetic peaks were fitted with a pseudo-Voigt function and integrated. In Fig. \ref{tempdep} (b) the temperature dependence of the integrated intensity of the magnetic reflection at (2,\,2/3,\,3) is shown. With increasing temperature magnetic order is suppressed and the intensity of the magnetic reflections decreases up to the N\'eel temperature $T_\mathrm{N}$. Above $T_\mathrm{N}$ only an experimental background signal is detected. From the data the transition temperature is determined to $T_\mathrm{N}\nobreak=\nobreak19.5$\,K. This is in good agreement with the reported N\'eel temperature $T_\mathrm{N}\nobreak=\nobreak20$\,K \cite{Mentink1994,Mentink1995,Mentink1996}.
	
	Furthermore, these data allow an analysis of the critical behavior of the magnetic transition. In the inset of Fig. \ref{tempdep} (b) a double logarithmic plot of the background corrected integrated intensity $I-I_\mathrm{bg}$ of the magnetic reflection at (2,\,2/3,\,3) vs. $(T_\mathrm{N}-T)/T_\mathrm{N}$ is shown and fitted with a power law
	\begin{align}
		\label{critbe}
		I~\propto~\biggl (\frac{T_\mathrm{N}-T}{T_\mathrm{N}}\biggr )^{2\beta}
	\end{align}
	in a range close to $T_\mathrm{N}$ ({\it i.e.}, from $16$\,K to $T_\mathrm{N}$). From this fit we obtain the critical exponent $\beta$, arriving for $\mathrm{UNi_4B}$ at a value $\beta \nobreak=\nobreak0.32(1)$. This would be in good agreement with the theoretically predicted value $\beta \nobreak=\nobreak0.3265(3)$ for a 3D-Ising like antiferromagnet \cite{Pelissetto2002}. The strong anisotropy of the susceptibility along and perpendicular to the hexagonal plane reported previously \cite{Mentink1994} supports the results that $\mathrm{UNi_4B}$ is an Ising-like system. Considering the noncollinear magnetic structure of UNi$_4$B, this results appears rather surprising. With its magnetic behavior and the anisotropy in the susceptibility, UNi$_4$B shows some similarities to a Dy$^{3+}$ molecular magnet that was proposed by Luzon et al. \cite{Luzon2008} to be an archetype of the noncollinear Ising model. Further investigations of the magnetic correlations and the in-plane magnetic anisotropy are necessary to establish if UNi$_4$B can be described by a similar noncollinear Ising model.
		
	Finally, we have carried out neutron diffraction experiments on $\mathrm{UNi_4B}$ single crystals in the temperature range of the proposed second ordering transition \cite{Movshovich1999}. If the assumption of a second magnetic transition would be correct, we would expect at temperatures below $330~\mathrm{mK}$ either additional magnetic reflections or a change of the magnetic intensity on the existing magnetic and/or nuclear reflections.

	First, we have checked for additional magnetic reflections with the CYCLOPS neutron Laue diffractometer of the ILL Grenoble. In Fig. \ref{cyclops} we display a section of the Laue pictures taken at temperatures of (a) $0.04$\,K, (b) $1.2$\,K and (c) $30$\,K. Above the Néel temperature $T_\mathrm{N}\nobreak=\nobreak19.5$\,K only the structural reflections are visible. With cooling below $T_\mathrm{N}$ additional magnetic reflections~are visible (indicated by arrows in Fig. \ref{cyclops}). Surprisingly, down to temperatures of $40~\mathrm{mK}$ and allowing in~the experiment for high statistics no additional magnetic reflections have been observed. 
	
	Next, using the D10 instrument, the temperature dependence of several magnetic and nuclear reflections has been measured down to temperatures of $100~\mathrm{mK}$. As shown in Fig. \ref{magn_low}, no resolvable change of the magnetic intensity was measured on any of these peaks at or below $330~\mathrm{mK}$. Therefore, at this point our findings are inconsistent with the proposal of a second magnetic phase transition at temperatures $\sim 330~\mathrm{mK}$. The anomaly measured in the specific heat in Ref. \cite{Movshovich1999} appears to have a different physical mechanism. The proposal of crystal field effects put forth in Ref. \cite{Yanagisawa2021} we cannot test with our elastic experiments.
	
	\section{Conclusion}
	
	Altogether, from our neutron diffraction study we conclude that $\mathrm{UNi_4B}$ is a rare example of an Ising-like frustrated metallic $f$-electron magnet with a partially ordered magnetic ground state. We have thoroughly characterized the crystal structure of UNi$_4$B, which turns out to be of lower symmetry than previously reported. Even on this lower-symmetry crystallographic lattice, we have verified a unique type of magnetic order, {\it i.e.}, a vortex-type spin structure of magnetic moments $\mu_{\mathrm{ord}}\nobreak=\nobreak 0.99(1)~\mu_\mathrm{B}$/(U atom) on 2/3rds of the U sites. The vortex structure carries in it the possibility of toroidal order existing in $\mathrm{UNi_4B}$, with a multitude of consequences for the observation of novel types of magnetoelectric effects. In addition, the refinement of the magnetic structure suggests that the magnetic moments are slightly canted compared to the perfect vortex structure. Finally, the assumption that there is another magnetically ordered phase at low temperatures could experimentally not be validated by means of our neutron diffraction experiments.
	
	\begin{acknowledgments}
		The authors are grateful to C. Tabata for fruitful discussions. We acknowledge the ILL for the allocation of neutron radiation beamtime. The complete sets of experimental data of individual experiments can be found in datasets \cite{ILL,ILL2,ILL3}. The UNi$_4$B crystal has been grown and characterized in MGML (https://mgml.eu/) which is supported within the program of Czech Research Infrastructures (Project No. LM2018096). The work of Czech co-authors was supported by MEYS CR (project No. LTT20014).
	\end{acknowledgments}


\begin{thebibliography}{00}
\bibitem{powder} The marginal difference between the refinements with the space groups \textit{Cmcm} and \textit{Pmm}2 can be explained with the low intensity (up to 0.1\,\% of maximum intensity) of the reflections forbidden in \textit{Cmcm}. For comparison, the background in the powder experiment was around 6\,\% of the maximum intensity. So we conclude that the forbidden reflections are not clearly distinguishable from the background and thus the differences in the crystal structures cannot be resolved in the powder experiment.
\end{thebibliography}
\end{document}